\documentclass[%
 reprint,
 amsmath,amssymb,
aps,
]{revtex4-2}

\usepackage{tikz}
\usepackage{graphicx}
\usepackage{dcolumn}
\usepackage{subcaption}
\usepackage{amsmath}
\usepackage{tcolorbox}
\usepackage{bm}
\usepackage{float}
\usepackage{multirow}
\usepackage[version=4]{mhchem}
\usepackage{xcolor}
\usepackage{enumitem}
\usepackage[version=4]{mhchem}
\usepackage{makecell}
\usepackage{bibunits}

\usepackage{booktabs}

\newcommand\Iiii{CsPbI$_3$}

\newcommand{\CP}{\ensuremath{\Delta \mu}}
\newcommand{\FE}{\ensuremath{\Delta H}}

\begin{document}

\preprint{APS/123-QED}
\title{Defect Tolerance and Local Structural Response to 3d Transition-Metal Substitution in \Iiii}
\author{Misbah Shaheen and Sheharyar Pervez}
 \email{sheharyar@giki.edu.pk}
\affiliation{
    Ghulam Ishaq Khan Institute of Engineering Sciences and Technology
}%

\date{\today}
\date{January 11, 2025}

    \begin{abstract}
        We present a systematic first-principles study of substitutional 3d transition-metal (TM) defects in \Iiii\ using the spin-polarized GGA+$U$ framework. TM incorporation is generally energetically favorable and induces lattice distortions that are strongly localized around the defect site, preserving the overall structural integrity of the host. Analysis of defect formation energies and electronic structure shows that, with the exception of Sc and Ti, \Iiii\ exhibits a strong resistance to deep trap formation. Most TM substitutions instead introduce resonant states that hybridize with the band edges, consistent with the defect-tolerant nature of the material. While these states can modify the band gap, they do not generate isolated mid-gap traps.
The observed distortions arise from strain-driven Van Vleck modes governed by ionic-radius mismatch, electronegativity differences, and TM-I orbital overlap, with amplitudes that decay rapidly away from the defect. Spin-polarized calculations reveal significant TM-induced spin polarization on the ligands and, in some cases, on neighboring Pb atoms, reflecting variations in covalency and hybridization across the 3d series. Together, these results establish a unified picture in which local structural response, electronic hybridization, and spin polarization jointly control the stability and electronic impact of TM defects in \Iiii, identifying dopants that are electronically benign or detrimental.
    \end{abstract}

    \maketitle

\begin{bibunit}[apsrev4-2]

    \section{\label{introduction}Introduction}

Efficiency of photovoltaic devices is linked to the appearance of the additional energy states due to the defects present in the system \cite{wang2024defects}. 
Shallow states introducing free carriers into the system, serves as the usual and conventional purpose of doping. 
Deep trap states located far from the band edges typically act as non-radiative recombination centers and are particularly detrimental because they strongly capture and immobilize carriers. 
Electrons in the conduction band or holes in the valence band may become localized in these states, leading to recombination events that do not contribute to useful current generation. 
Such processes reduce the minority carrier lifetime, decrease the open-circuit voltage ($V_{\text{oc}}$), and limit the overall power conversion efficiency (PCE) of the solar cell 
\cite{motti2019defect}\cite{ahn2016trapped}\cite{zhang2019ab}\cite{spectroelectrochemical}\cite{ali2024metal}. 
On the other hand, 
defects whose states lie outside the band gap or are successfully passivated may render the material more robust against electronic degradation. 
For instance, resonant states are typically fully occupied (in the valence band) or empty (in the conduction band), and therefore do not function as active carrier traps\cite{biswas2025nature}. 
In this case, the defect is electronically benign and can even contribute to thermodynamic stability, as the defect states hybridize with bulk states rather than forming localized gap states \cite{pandey2016band}\cite{das2021defect}\cite{zhang2023insight}. 
Similarly, passivation of trap states, for example, through hydrogenation or halide additives, can stabilize the material and reduce recombination losses \cite{wu2024defect}.

    In this context, inorganic halide perovskites (IHPs) are widely regarded as defect-tolerant materials, allowing targeted modification of specific properties while leaving the overall electronic structure largely intact \cite{kang2017high}. Experimental studies have provided valuable insights into how defects influence the macroscopic properties of IHPs \cite{yin2014unique,dobrovolsky2017defect,bao2022physics}. 
    This high defect tolerance has been attributed to the predominantly shallow nature of native defects, which limits carrier trapping \cite{Self-regulation}. Complementary theoretical work has systematically examined defect formation energies and charge-state stability in CsPbX$_3$ compounds, revealing a pronounced halide dependence of defect energetics that strongly affects carrier trapping behavior and material stability \cite{buin2014materials}.

    Despite these advances, establishing a complete and unified understanding from empirical observations remains challenging. In particular, identifying the exact types of defects present under varying synthesis and operating conditions \cite{Imperfections}, and correlating them reliably with their associated trap states \cite{Thin-film}, continues to be an open problem.

    Meanwhile, computational studies of point defects have evolved into a mature and predictive framework, driven by advances in first-principles methods and computational resources. While early density functional theory (DFT) based defect calculations in the 1990s established the basic formalism, the field has developed rapidly over the last decade. Improvements in exchange-correlation treatments, finite-size correction schemes, and automated workflows have enabled reliable prediction of defect formation energies, charge transition levels, and carrier trapping behavior across complex materials classes. As a result, defect calculations now play a central role in understanding defect tolerance, recombination processes, and dopant behavior in functional materials.

In this study, we perform a systematic first-principles based investigation of substitutional 3d transition-metal defects in \Iiii\ using spin-polarized GGA+$U$ calculations. 
We examine defect formation energetics, local structural distortions, and the resulting electronic and magnetic responses across the full 3d series. 
By linking localized lattice distortions with electronic hybridization and spin polarization effects, we develop a consistent physical picture of how resonant and deep defect states emerge. 
Our analysis demonstrates that the lattice response to TM substitution is highly localized and that \Iiii\ is intrinsically resistant to deep trap formation, with only a limited subset of dopants inducing electronically detrimental states. 
These results provide a unified framework for understanding transition-metal defect tolerance in \Iiii\ and offer clear guidance for identifying benign and harmful dopants in halide perovskites.

    \section{\label{preliminaries}Preliminaries}

Dopants induce defects in semiconductors and photovoltaic (PV) materials 
which create new electronic states within the electronic band structure.
These defects, broadly classified in Table. \ref{defecttype}, may be
\textit{deep/trap} (inside bandgap far from edges), 
\textit{shallow} (inside bandgap near the valence or conduction band) or 
\textit{resonant states} (inside VBM or CBM).
Overall, trap states are generally detrimental to photovoltaic efficiency because they act as recombination centers that limit carrier lifetime and voltage.   
Although intermediate-band solar cells aim deliberately engineered trap-like states that allow absorption of sub-bandgap photons, potentially enhancing photocurrent; achieving this effect without introducing significant recombination losses remains a challenge. 
A careful understanding of both the energetic position and the chemical passivation of trap states is thus central to the design of next-generation, stable, and efficient photovoltaic devices. 

\begin{table}
\centering
\caption{\label{defecttype} Classification of defect states by type and energetic position.}
\renewcommand{\arraystretch}{0.9}
\begin{tabular}{lcccc}
\toprule
\makecell{Defect} &
\makecell{Inside\\bandgap} &
\makecell{Merges\\with\\VBM/CBM} &
\makecell{Carrier trapping\\and\\recombination} &
\makecell{Acceptors\\or\\Donors} \\
\midrule
Resonant & $\times$& \checkmark  & $\times$ & \checkmark \\
Deep     & \checkmark& $\times$  & \checkmark & $\times$ \\
Shallow  &  \checkmark& $\times$ & $\times$ & \checkmark \\
\bottomrule
\end{tabular}
\end{table}



\emph{Defect Formation Energies (DFEs) ---}
Thermodynamically, DFEs describe the ease with which a certain defect can be formed. For the case of a defect \textit{X} with charge \textit{q} in a host structure, this can be written as:


\begin{align}
    \Delta H(X,q) = &E_{\text{X}} - E_{\text{host}} - \sum_i n_i \mu_i \nonumber \\
    &+ q[E_{VBM_{\text{host}}} + E_F + \Delta V] + E_{\text{corr}}
    \label{eq_DFE}
\end{align}

Where $E_{\text{X}}$ and $E_{\text{host}}$ represent the total energy of defected and pristine structures, respectively. 
$n_i$ is the number of atoms added ($n_i > 0$) or removed ($n_i < 0$) from the system for which the energy $\mu_i$ is required. 
The charge of the defect $q$ is the number of electrons transferred to or from the reservoir to create a defect, 
$E_{\text{VBM}}$ is the valence band maximum (VBM) of the host material, and
$E_F$ is the Fermi level which defines the electrochemical potential of the electrons (add or remove) to create a charged system with charge $q$. 
$E_{\text{corr}}$ is the energy correction.
Determining the VBM in defective systems is tricky because defects perturb the potential energy surface by introducing states inside the bandgap. Thus the VBM of the pristine strucutre is usually taken as the reference point 
, albeit shifted by a potential alignment term $\Delta V$, calculated by comparing the core level energy of pristine and defective systems. 
$E_F$ then is the energy added on top of the reference point $E_{\text{VBM}} + \Delta V$ and usually taken as zero at VBM. 

\emph{Energy Corrections ---} 
DFT codes frequently employ periodic boundary conditions which mandates the use of the supercell method to isolate the defect and prevent the interactions between periodic images. For charged defects, the long range nature of the Coulomb interaction then yields divergent electrostatic behavior. 
To avoid using a very large supercell, $E_{\text{corr}}$ is the correction needed to eliminate the effects of electrostatic interactions between charged defects in periodic images \cite{freysoldt2014first} \cite{naik2018coffee}. \\
The charge correction scheme used here, was introduced by Freysoldt, Neugebauer, and Van de Walle (FNV) \cite{freysoldt2014first} and requires the alignment of the potentials of charged and neutral defected supercells due to the undefined reference electrostatic potential. 
The term $E_{\text{corr}}$ in eq. \ref{eq_DFE} then takes the general form:

\begin{equation}
    E_{\text{corr}} = E_q^{lat} - q \Delta_{q/0}
    \label{eq-Ecorr}
\end{equation}

where $E_q^{lat}$, the macroscopically screened lattice energy of defect charge with compensating background, accounts for the long range interactions and $q \Delta_{q/0}$ is the alignment term.\\

\emph{Charge State Transition Levels (CSTLs) ---}
Charge state transitions take place when the DFEs for two different charge states $q$ and $q'$ become equal.
The Fermi energies corresponding to these points are the CSTLs. CSTLs are a useful indicator of whether the new energy state is a detrimental trap or a useful energy state and are calculated using \cite{zhang2008dopant}:

\begin{equation}
    \epsilon (q/q') = \frac{\Delta H (X^q, E_F = 0) - \Delta H(X^{q'}, E_F =0)}{q'-q}
    \label{eq_CSTL}
\end{equation}

Here, $\Delta H$ shows the DFE of charge state $q$ or $q'$ at VBM i.e. $E_F = 0$. \\

\begin{figure}
    \includegraphics[scale=0.21]{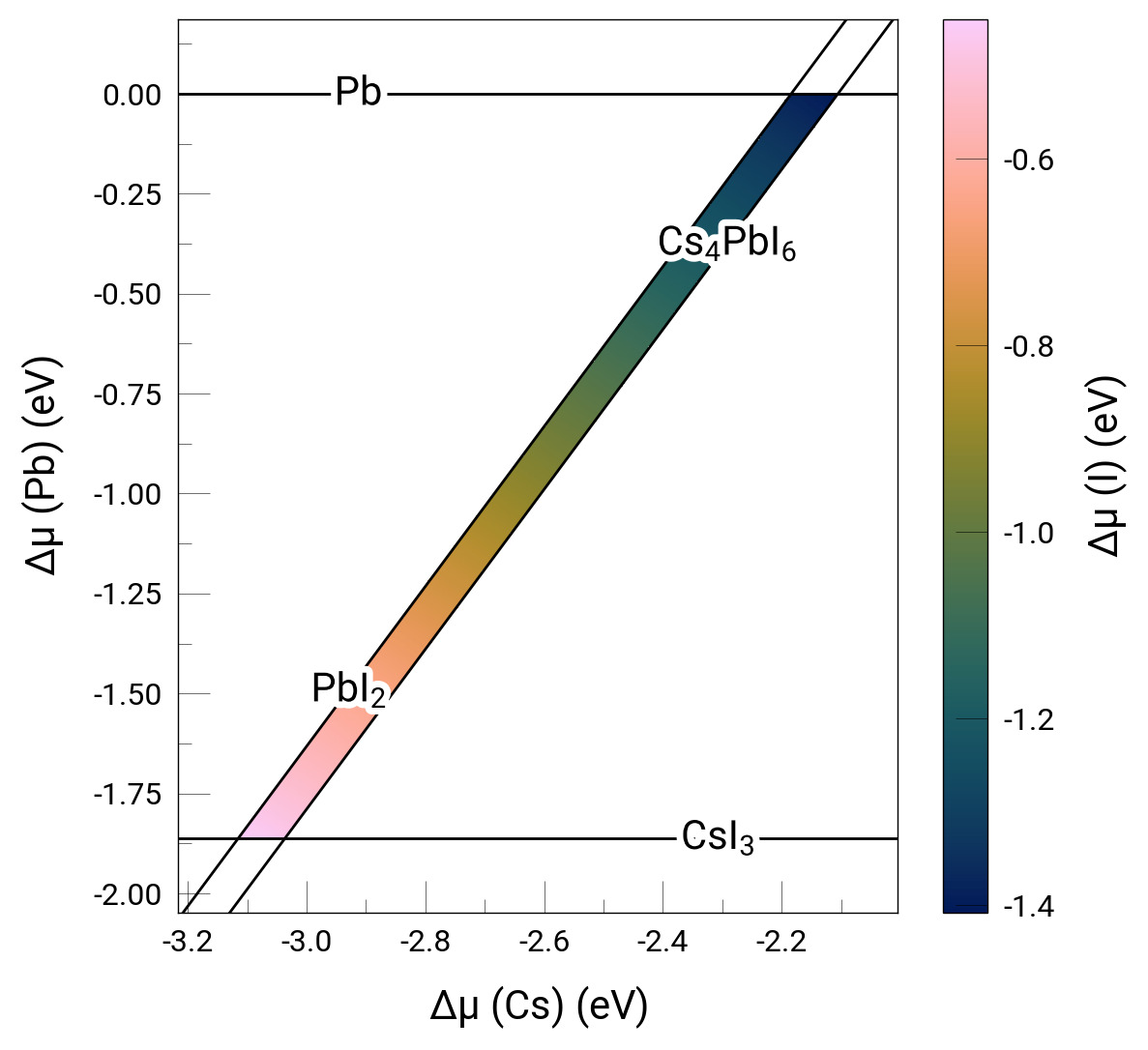}
    \caption{Stability polygon showing the stable region for \Iiii\ enclosed by competing phases.}
    \label{chempots}
\end{figure}

\renewcommand{\arraystretch}{1.8}
\begin{table*}
\caption{\label{t1_chempots}Chemical potentials for elemental species present in \Iiii\ in equilibrium with $PbI_2$ and $Cs_4PbI_6$ for I-rich and I-poor environmental conditions}
\begin{ruledtabular}
\begin{tabular}{lcc}

 & I-rich & I-poor \\
\hline

\multirow{3}{*}{\parbox{2.0cm}{Equillibrium with $PbI_2$}} & $\CP (I) = \frac{1}{2}\CP (I_2^{mol})$ & $\CP (I) = \frac{1}{2}[\CP (PI_2) - \CP(Pb^{bulk})]$\\
    & $\CP(Pb) = \CP(PbI_2) -\CP(I_2^{mol}) $ & $\CP (Pb) = \CP (Pb^{bulk})$\\
    & $\CP(Cs) = \FE(CsPbI_3) - \CP (PbI_2) - \frac{1}{2} \CP (I_2^{mol})$ & $\CP(Cs) = \FE(CsPbI_3) + \frac{1}{2}\CP(Pb^{bulk} - \frac{3}{2}\CP(PbI_2))$\\

\addlinespace[5mm]

\multirow{3}{*}{\parbox{2.0cm}{Equillibrium with $Cs_4PbI_6$}} & $\Delta \mu (I) = \frac{1}{2}\Delta \mu (I_2^{mol})$ & $\CP(I) = \frac{5}{3} \CP(CsPbI_3) - \CP(Pb^{bulk}) - \frac{1}{3} \CP(Cs_4PbI_6)$\\
&  $\CP(Pb) = \frac{4}{3}\FE(CsPbI_3) - \frac{1}{3}\FE(Cs_4PbI_6) -  \CP(I_2^{mol})$ & $\CP (Pb) = \CP (Pb^{bulk})$\\
& $\CP(Cs) = \frac{1}{3} [\FE(Cs_4PbI_6) -\FE(CsPbI_3)]  - \frac{1}{2} \CP(I_2^{mol})$ & $\CP(Cs) = \frac{1}{3} \FE(Cs_4PbI_6) - \frac{2}{3} \FE(CsPbI_3)$ \\

\end{tabular}
\end{ruledtabular}
\end{table*}

\emph{Phase stability ---}
The range of chemical potentials (CPs) is thermodynamically constrained by the competing phases
surrounding the region in which the host material \Iiii\ is the most stable phase. These constraints depend on the growth conditions and must be determined by keeping the stable region of host material in view. 
Thus the elemental chemical potentials are constrained by the formation enthalpy $\Delta H$ of \Iiii, such that:

\begin{equation}
    \Delta \mu (Cs) + \Delta \mu(Pb) + 3 \Delta \mu(I) = \Delta H (CsPbI_3)
    \label{eq_hoststability}
\end{equation}

where $\Delta \mu$ are the effective atomic CPs of the associated element in the perovskite environment with reference to the standard elemental phase. 
The remaining constraints prevent the formation of secondary phases (PbI$_2$, Cs$_4$PbI$_6$, CsI$_3$, and Pb) are:

\begin{align}
\begin{split}
    \Delta \mu (Pb) + 2 \Delta \mu (I) &< \Delta H (PbI_2) \\
    4 \Delta \mu (Cs) + \Delta \mu (Pb) + 6 \Delta \mu (I) &< \Delta H (Cs_4PbI_6)\\
    4 \Delta \mu (Cs) + 3 \Delta \mu (I) &< \Delta H (CsI_3)\\
    \Delta \mu (Pb) &< 0
    \label{eq_CPs}
\end{split}
\end{align}

Eq. \ref{eq_CPs} defines the stability polygon shown in Fig. \ref{chempots}. The conditions on CPs to obtain stable \Iiii in equilibrium with PBI$_2$ and Cs$_4$PbI$_6$ in I-rich and I-poor environment are given in Table \ref{t1_chempots}.

\emph{Octahedral Dsitortions ---} Displacement of ligands coordianted with the central atom distorts the octahedron and can be described through Van Vleck modes \cite{van1939jahn}\cite{rondinelli2011structure}. 

Bond-length distortion modes relevant to this study are (i) the isotropic breathing mode $Q_1$, which describes the uniform expansion/contraction of the octahedron, and (ii) the tetragonal mode $Q_3$, where axial and equatorial ligands behave differently, resulting in a compressed or elongated octahedron. 
The magnitude of distortion, $\rho_0$, for the $Q_3$ mode is given by:

\begin{equation}
    \rho_0 = \sqrt{Q_2^2 + Q_3^2}
    \label{dist_magnitude}
\end{equation}

where the magnitude is determined using the basis vectors $Q_2$ and $Q_3$.

Bond-length distortion index is defined by Baur\cite{baur1974geometry} as:

\begin{equation}
    D = \frac{1}{n} \sum_{i=1}^n \frac{|l_i - l_{avg}|}{l_{avg}}
    \label{bondlength_distortion}
\end{equation}

where $n$ is the number of ligands, $l_i$ being the distance between central atom and the $i$-th ligand, and $l_{avg}$ is the average of all distances between central atom and ligands. 

Effective coordination number (ECN) \cite{hoppe1979effective} is another parameter which describes the deviation from the regular octahedral geometry:

\begin{equation}
    ECN = \sum_{i=1}^n exp 
    \Biggl [ 
        1-\biggl ( 
            \frac{l_i}{l'_{avg}}{)}^6
    \Biggr ]
    \label{ECN}
\end{equation}

Here, $l'_{avg}$ is the modified average of distances which is equal to:

\begin{equation}
    l'_{avg} = \frac{\sum_{i=1}^n l_i exp[1-(l_i/l_{min})^6]}{\sum_{i=1}^n l_i exp[1-(l_i/l_{min})^6]}
    \label{modified_avg_dist}
\end{equation}
    \section{\label{methodology}Methodology}

\begin{figure*}
    \includegraphics[scale=0.5]{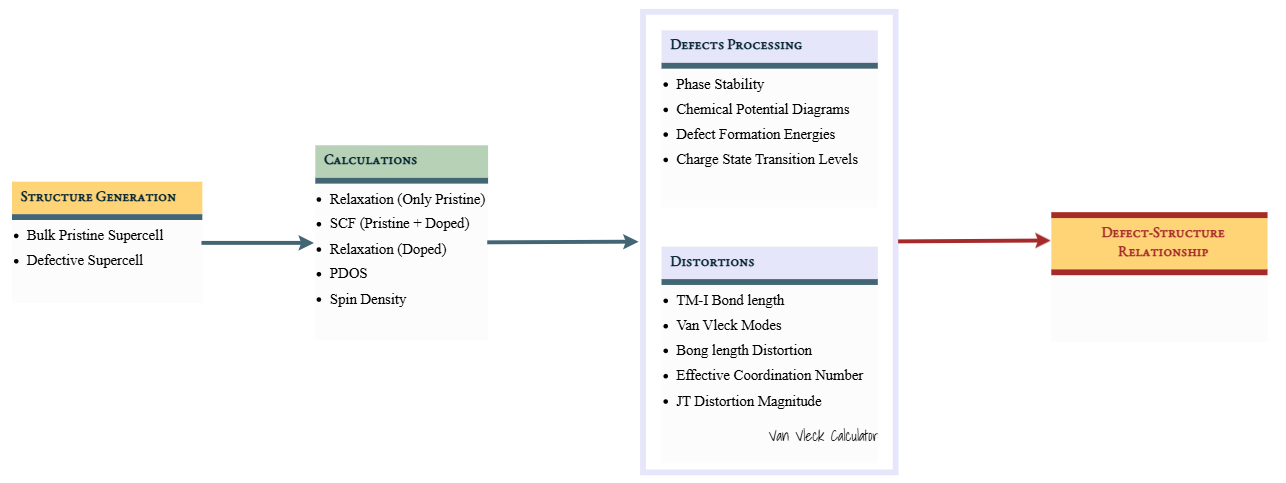}
    \caption{\label{workflow}Workflow demonstrates the methodology adapted for this study.}
\end{figure*}

Fig. \ref{workflow} shows the workflow adopted for this study. Detailed explanation of each step is given below:

\emph{Structure Generation ---} Defected structures were generated using \texttt{VESTA} \cite{VESTA} by taking a $2\times2\times2$ supercell of the host material, \Iiii\, and replacing Pb at the origin with 3d-TMs (Sc, Ti, V, Cr, Mn, Fe, Co, Ni, Cu, Zn) one by one as point defects in the crystal.

Goldschmidt tolerance factor (t) \cite{goldschmidt_gesetze_1926} octahedral factor ($\mu$) were calculated using the equations below:

\begin{equation}
    t = \frac {r_A + r_X} {\sqrt{2} (r_{B_{eff}} + r_X)}
    \label{eq_t}
\end{equation} 

\begin{equation}
    \mu = \frac{r_{B_{eff}}}{r_X}
    \label{eq_mu}
\end{equation} 

where, $r_A$ is the radius of Cs, $r_{B_{eff}}$ is the effective radius of atoms on B-site including Pb and the defect, and $r_X$ is the I radius. Effective radius for B-site is equal to $(7\times r_B + r_{defect})/8.0$.

\emph{DFT Calculations ---} DFT calculations were performed using Quantum Espresso v.7.4 (QE) \cite{QE-2009} with GGA-PBE functionals to optimize the geometry and obtain energies of bulk and defected structures. The host supercell was fully relaxed as bulk, after which the defected supercells were created using relaxed coordinates of the host material. 
The Hubbard correction (See Supplementary Table \ref{hubbard}) was implemented for the transition metals within the spin-polarized GGA+$U$ framework to obtain accurate results.
Relaxations, partial density of states (PDOS), and spin density calculations of doped structures was also performed for further analysis. 

Materials project \cite{MP} and python libraries including \texttt{pymatgen} \cite{pymatgen} and \texttt{doped} \cite{kavanagh2024doped} were used for defect calculations to obtain DFE (eq. \ref{eq_DFE}), CSTLs (\ref{eq_CSTL}), competing phases, Freysoldt energy corrections etc. The functionality of \texttt{doped} was expanded to add support for QE and is available on github. 

Relaxed doped structures were further analyzed using the python package \texttt{VanVleck Calculator} \cite{nagle2024van} and \texttt{VESTA} to determine the distortion parameters as mentioned in Fig. \ref{workflow}.

    \section{\label{resultsanddiscussion}Results and Discussion}

\subsection*{Defect Analysis}

\begin{table}
\caption{\label{structuralproperties}Structural properties of defected structures.}
\renewcommand{\arraystretch}{0.9}
\begin{ruledtabular}
\begin{tabular}{lrrrr}
 
 Material & \makecell[r]{Oxidation\\ state}  & \makecell[r]{Ionic \\radius} & \makecell[r]{Goldschmidt \\ factor ($t$)} & \makecell[r]{Octahedral\\ factor ($\mu$)}  \\
         \hline
         \vspace{-1mm}&&\\
         \multirow{1}{*}{\parbox{2.2cm}{\Iiii}}  & -  & - & 0.85 & 0.54 \\
         \addlinespace
         \multirow{2}{*}{\parbox{2.2cm}{$\ce{Sc_{Pb}}$}}& +2 & 0.83&0.86& 0.52 \\
                                              & +3 & 0.74 & 0.87&0.52 \\
                                              \addlinespace
         \multirow{3}{*}{\parbox{2.2cm}{$\ce{Ti_{Pb}}$}}& +2& 0.86 & 0.86&0.52 \\
                                      & +3 & 0.67 & 0.87&0.51 \\
                                      & +4 & 0.61 & 0.87&0.51 \\
                                      \addlinespace
         \multirow{3}{*}{\parbox{2.2cm}{$\ce{V_{Pb}}$}}& +2& 0.79 & 0.86&0.52 \\
                                      & +3 & 0.64 & 0.87&0.51 \\
                                      & +4 & 0.58 & 0.87&0.51 \\
                                      & +5 & 0.54 & 0.87&0.50 \\
                                      \addlinespace
         \multirow{3}{*}{\parbox{2.2cm}{$\ce{Cr_{Pb}}$}}& +2& 0.8 & 0.86&0.52 \\
                                      & +3 & 0.6 & 0.87&0.51 \\
                                      & +6 & 0.44 & 0.88&0.49 \\
                                      \addlinespace
         \multirow{2}{*}{\parbox{2.2cm}{$\ce{Mn_{Pb}}$}}& +2& 0.83 & 0.86&0.52 \\
                                      & +4 & 0.53 & 0.87&0.50 \\
                                      & +7 & 0.46 & 0.87&0.49 \\
                                      \addlinespace
         \multirow{3}{*}{\parbox{2.2cm}{$\ce{Fe_{Pb}}$}}& +2& 0.78 & 0.86&0.52 \\
                                      & +3 & 0.65 & 0.87&0.51 \\
                                      \addlinespace
         \multirow{2}{*}{\parbox{2.2cm}{$\ce{Co_{Pb}}$}}& +2& 0.745& 0.87&0.52 \\
                                      & +3 & 0.61 & 0.87&0.51 \\
                                      \addlinespace
         \multirow{2}{*}{\parbox{2.2cm}{$\ce{Ni_{Pb}}$}}& +2& 0.69 & 0.87&0.51 \\
                                      & +3 & 0.6 & 0.87&0.50 \\
                                      \addlinespace
         \multirow{2}{*}{\parbox{2.2cm}{$\ce{Cu_{Pb}}$}}& +2& 0.73 & 0.87&0.51 \\
                                      & +1 & 0.77 & 0.86&0.52 \\
                                      \addlinespace
         \multirow{1}{*}{\parbox{2.2cm}{$\ce{Zn_{Pb}}$}}& +2& 0.74 & 0.87&0.51 \\

\end{tabular}
\end{ruledtabular}
\end{table}

\begin{figure}
    \includegraphics[width=3.5in]{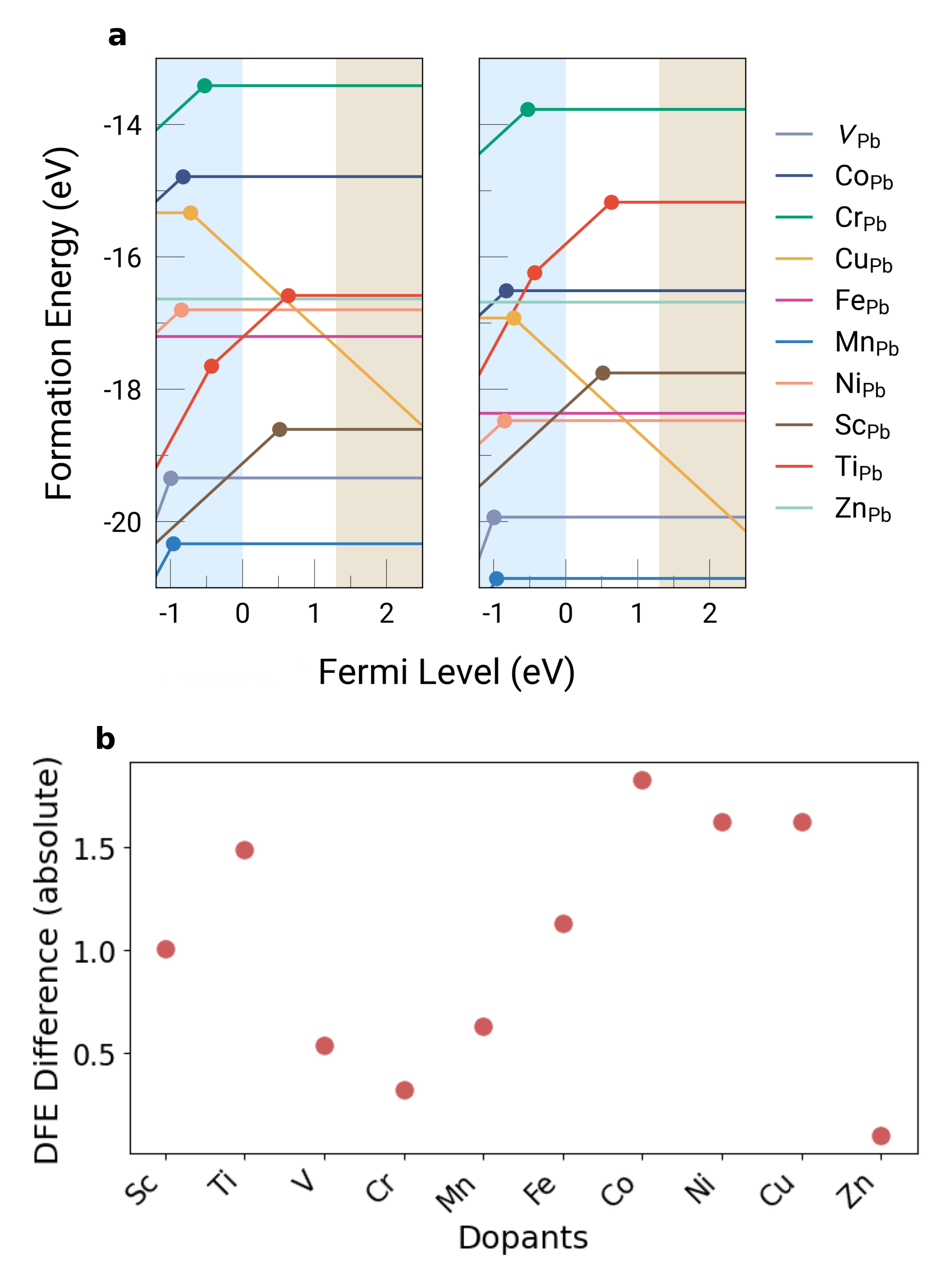}
    \caption{(a): Defect formation energies in I-poor (left) and I-rich (right) shows Mn is the most stable defect with lowest formation energy. (b) Absolute difference of DFEs in I-rich and I-poor environment indicating dependence of defects on growth conditions.}
    \label{DFEs}
\end{figure}

Before moving to the DFE analysis, a comparison of the Goldschmidt tolerance (\(t\)) and octahedral factors $\mu$ of the defected structures (Table \ref{structuralproperties}) with ionic radii (Supplementary Fig. \ref{tmuvsradii}) shows a linear relationship. All defects exhibit a slight increase over the \(t=0.85\) value of the pristine structure, accompanied with modest octahedral distortion. Nonetheless, all $\mu$ values remain within the usually accepted stability range for perovskites\cite{tmurange}. The tolerance factor increases linearly with decreasing ionic radius. When grouped by oxidation states, the DFE falls exponentially with Shannon ionic radii (Table \ref{structuralproperties}) (Supplementary Fig. \ref{DFEvsionicradii}) with the 0.69 \AA being the smallest radius corresponding to a stable defect of Ni.

To assess the stability of defects we calculate the DFEs and charge state transition levels (CSTLs), of the defected \Iiii\ systems in their neutral as well as physically possible charged states. The DFEs for all defects at the VBM are reported in Table \ref{table_Ipoor} and \ref{table_Irich}, and plotted as a function of the Fermi energy in Fig. ~\ref{DFEs}(a). 

Most defects have a relative charge of zero in the bandgap corresponding to horizontal lines. Mn and Cr are the most and least stable defects in both environments. At the I-poor end, Mn is followed by V and then Sc, but V, Fe, and Ni at the I-rich end. Close to the conduction band, all defects are neutral (in +2 oxidation state) with the exception of Cu which is in the +1 state across the entirety of the bandgap and is sloped downward because of its total relative charge of -1. Sc and Ti are the only ones that show transition levels in the bandgap going from +3 to +2. These transition levels create deep states which can act as recombination centers which can be detrimental to optical performance. Resonant states can be beneficial when causing no harm to electronic structure but they can also be detrimental by modulating the band edges as in the case of Cr (as explained later in density of states).

\begin{table}
\caption{\label{stabilityregion}Growth conditions in which defect is more stable.}
\renewcommand{\arraystretch}{0.9} 
\begin{ruledtabular}
\begin{tabular}{lll}
Stability region & Dopant & Oxidation states\\
\hline
\vspace{-1mm}&&\\
\multirow[t]{8}{*}{I-rich} & $\ce{V}$  & +2, +3, +4, +5\\
                                           & $\ce{Cr}$ & +2, +3, +6\\
                                           & $\ce{Mn}$ & +2, +4, +7\\
                                           & $\ce{Fe}$ & +2, +3\\
\addlinespace[6pt]  

                                           & $\ce{Co}$ & +2, +3\\
                                           & $\ce{Ni}$ & +2, +3\\
                                           & $\ce{Cu}$ & +1, +2\\
                                           & $\ce{Zn}$ & +2\\

\addlinespace[5pt] 

\multirow[t]{2}{*}{I-poor} & $\ce{Sc_{Pb}}$ & +2, +3\\
                                           & $\ce{Ti_{Pb}}$ & +2, +3, +4\\

\end{tabular}
\end{ruledtabular}
\end{table}

The stability of a defect with respect to its chemical environment is dependent on its growth conditions. Defects involving substitutions directly couple to chemical potentials and induce significant local relaxations, making their stability sensitive to growth conditions. Table \ref{stabilityregion} list the stable regions (I-rich or I-poor) for each defect whcih shows that with the exception Sc and Ti, which are more stable in I-poor region, all defects are more stable in I-rich region.

Fig. ~\ref{DFEs}(b) shows the absolute difference of DFEs (each defect with oxidation state corresponding to its lower DFE) I-rich and I-poor region. Sc, Ti, Fe, Co, Ni, and Cu are comparatively more sensitive to $\mu_I$ than Zn and Cr. The latter two are relatively independent of growth conditions, or put another way, these defects can always be easily created irrespective of the environment. For environment controlled defects (large differences), the choice of I-rich vs I-poor completely changes defect stability. Defects such as Mn and V, are moderately dependent on $\mu_I$ and can be stabilized or suppressed by cotrolling the growth conditions.

\subsection*{\label{relaxation}Stability Analysis}

Relaxation of materials in their stable oxidation states reveals the impact on bond lengths of both the dopant (TM) and non-dopant (Pb) octahedra as well as the extent of octahedral distortion itself (see Supplementary Table \ref{bondlengths}). The octahedral bond length 3.15 \AA for \Iiii\ decreases for the dopant-I with the extent of shortening depending upon various parameters such as ionic radii, orbital overlap, competition between different energies etc.

\begin{figure}
    \includegraphics[scale=0.6]{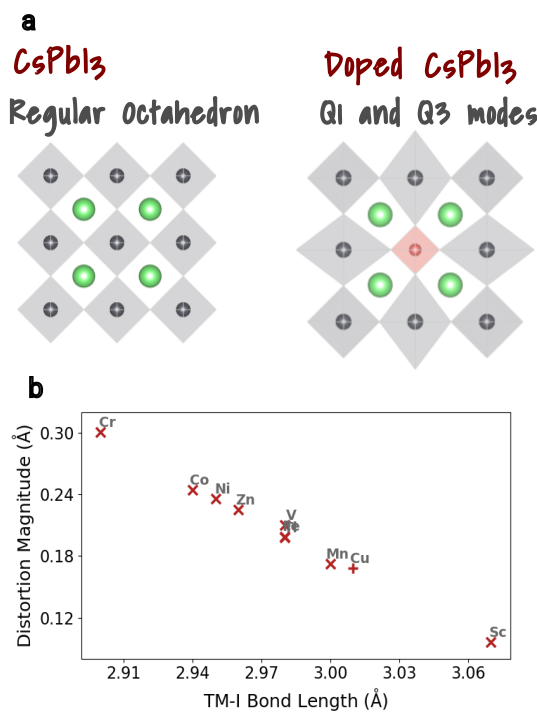}
    \caption{Strain driven dsitortions in local environment. (a) Left: Schematic showing the regular octahedrons in pristine \Iiii\; Right: Shrinked TM octahedron and distorted PbI$_6$. (b) Distortion magnitude of the PbI$_6$ octahedron adjacent to TM octahedron. Markers x and + indicate +2 and +1 oxidation states, respectively.}
    \label{octadist}
\end{figure}

While the dopant octahedra shrinks symmetrically, due to the smaller ionic radii, partially filled d-orbitals, and stronger electron sharing with p-orbitals, the adjacent Pb octahedraon is stretched asymmetrically. 
As illustrated in Fig. \ref{octadist}(a), this distortion in the local environment stems from the strain caused by the Van Vleck modes Q1 and Q3 for the TM and its adjacent PbI$_6$ octahedron respectively. 
The magnitude of the distortions decreases with increasing distance from the dopant site, indicating that the lattice accommodates the defect locally, without causing significant structural perturbations in the more distant regions (Fig. \ref{octadist}). 
Distortion parameters including distortion magnitude (Fig. \ref{octadist}), bond-length distortion index, and ECN were obtained (Supplementary Fig. \ref{distparams}) for the PbI$_6$ octahedron adjacent to TM octahedron. These distortion paramters have linear relation with TM-I bondlength which is the evidence of strain driven local distortion. Therefore, maximum distorted structure is obtained for Cr attributed to its shorer TM-I bond length.

\begin{figure*}
    \includegraphics[scale=0.9]{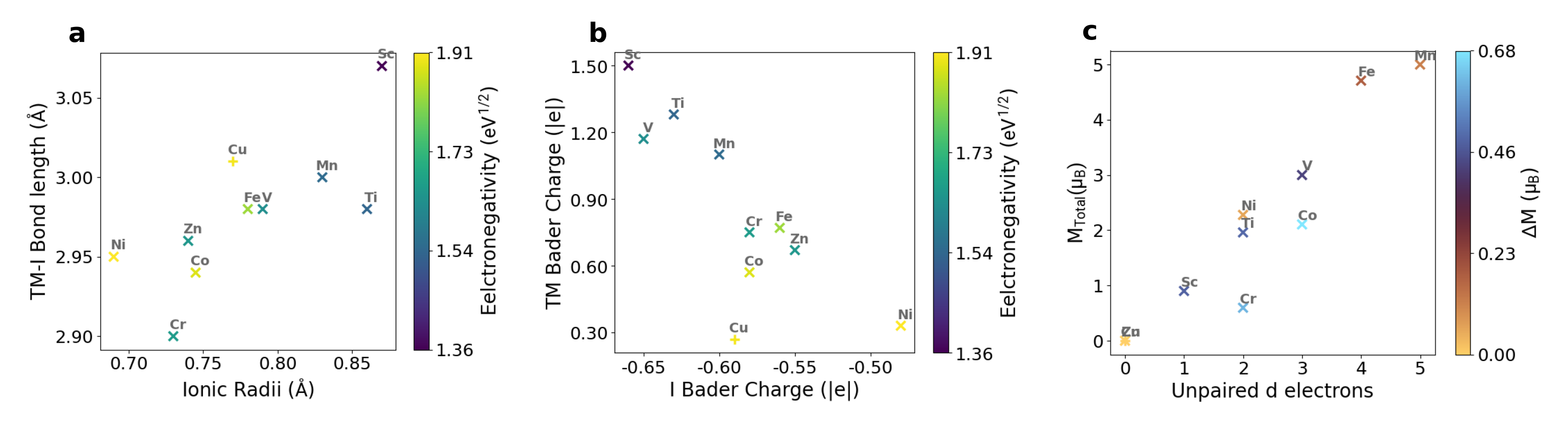}
    \caption{Trends and anomalies observed for defected perovskites. (a) TM-I bond length vs ionic radii showing a deviation from linear trend. (b) Bader charge analysis giving insights into bonding nature between TM and I. (d) Increase in magnetization with number of unpaired electrons with Cr as exception where colorbar $\Delta$ magnetization is the difference between total and absolute magnetizaiton. Markers x and + indicate +2 and +1 oxidation states, respectively.}
    \label{trends}
\end{figure*}


Fig. \ref{trends}(a) illustrates the relationship between TM ionic radii and the TM-I bond length along with the contribution of electronegativity ($\chi$). 

Large (small) ionic radii are expected to correlate with lower (higher) electronegativities and thus longer (shorter) bonds. Violating this trend are Zn and Cr whose electronegativities (1.65 and 1.66) are in the middle but create shorter bonds. With the exception of Sc and Cr, all bond lenghts lie in a window of roughly $0.7 \AA$.

Further light is shed by
 Bader charge analysis into the ionic vs. covalent nature of the bond between TM and I (Fig. \ref{trends}(b)). Large positive Bader charge on the TM and more negative Bader charge on the halide, both shows a more ionic behavior.
With I being highly polarizable, replacing Pb with a TM forces the TM to stabilize a relatively soft lattice. 
Early TMs (Sc, Ti, V, etc.) with their high tendency to oxidize result in a larger transfer of charge to I, than late TMs (Co, Ni, Cu) which resist oxidation.
Anomalies to this expected trend usually arise from either crystal field stabilization, preferred oxidation states, or spin-state changes.
    Mn with its half-filled high-spin configuration is exceptionally stable. On the other hand, Cr2+ (d$^4$) is notoriously unstable, tends to undergo more local symmetry breaking and electronic instability. Thus the Cr point is significantly off its expected position.
    Cu1+ (d$^10$) is very stable resulting in the lattice pushing charge onto I.
 
Fig. \ref{trends}(c) show the total magnetization as a function of the number of unpaired d-electrons of the dopant with the color bar showing the difference between the total and absolute magnetization. The total magnetization is the sum of all spin contributions but the absolute magnetization measures the magnitude of local moments and their difference indicates how much spin cancellation exists. TM in the octahedral environment of iodine (weak field ligand) are expected to achieve high spin configuration. Apart from Cu and Zn (which are non-magnetic), Mn, Fe, and Ni have minimum magnetization difference, indicating absence of spin compensation. V and Co, both have the 3 unpaired electrons but yield different total magnetizations. The magnitude of magentization difference in Co shows the existence of opposite spins. 
Cr has the lowest magnetization among all despite having 4 d electrons, possibly due to its low spin configuration in the system. 


\begin{figure*}
    \includegraphics[scale=0.8]{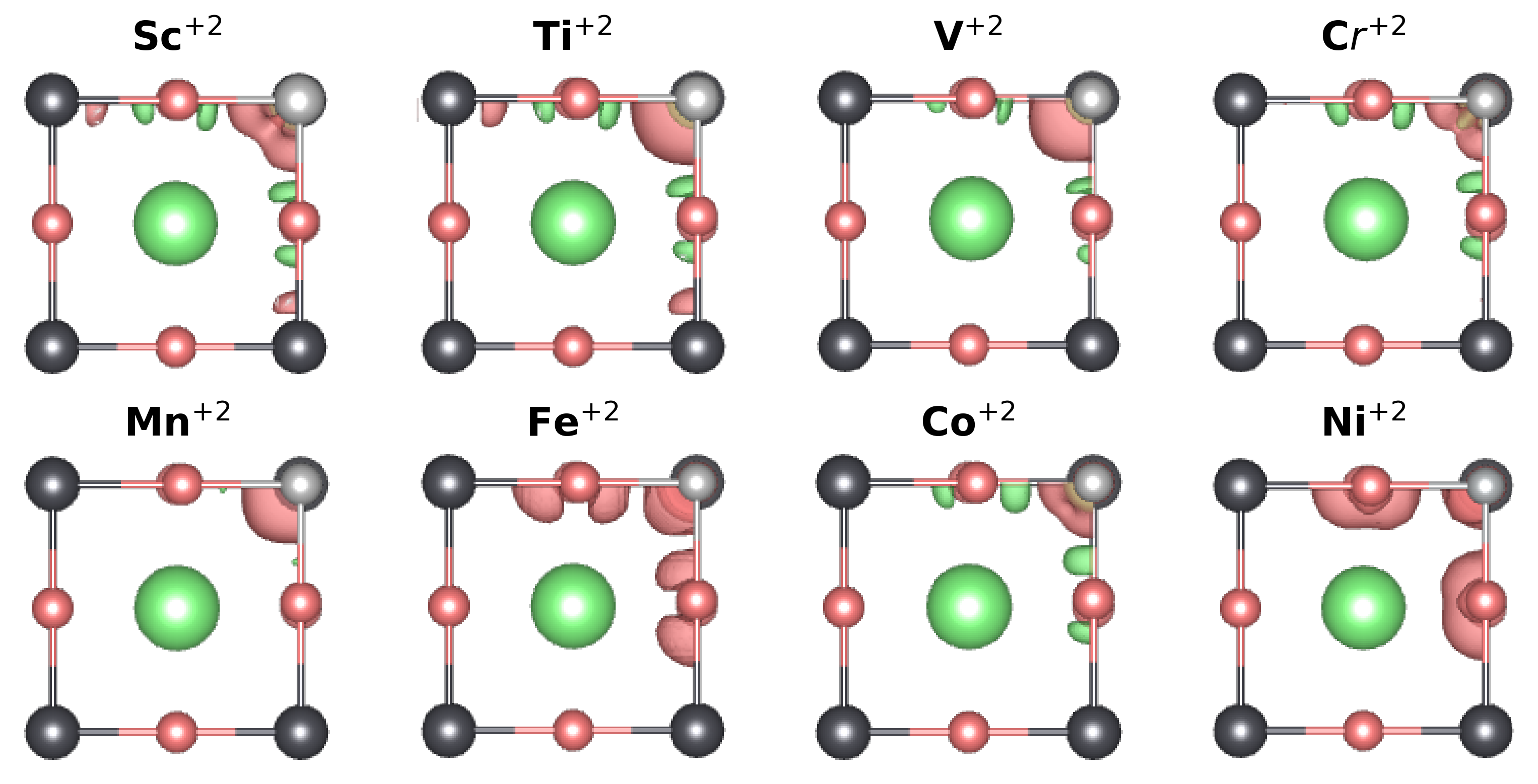}
    \caption{Spin densities; pink: positive and cyan: negative density with atoms: silver: TM, pink: I, dark gray: Pb and green: Cs. Mn has highly localzied density whereas other TMs induce spin density on nearby atoms.}
    \label{spindensities}
\end{figure*}

The results of magnetization and Bader charges correlate with the spin densities ($\Delta \rho = \rho_{\uparrow} - \rho_{\downarrow}$) in Fig. \ref{spindensities} (at isovalue of 0.001 \(eV/\AA^3\), with yellow for spin up and cyan for spin down) which were calculated to show the extent of spin polarization introduced by the hybridization of transition metal d-orbitals with I p-orbitals. 

All systems (except Cu and Zn) show polarization with the positive density centered on the dopant and spin leakage to the nearby I atoms.

The behavior of spin densities can be grouped into four bins:

\begin{enumerate}[label=\roman*)]
    \item In the case of Sc and Ti the spin leakage extends to the Pb atoms in the neighboring octahedron. The nearby I ligands are polarized with opposite spins (i.e. negative spin density) whereas the Pb atoms further down the line are again polarized with up spins. 
    This alternating polarization scheme (TM $\uparrow$, I$\downarrow$, Pb$\uparrow$) suggests the existence of superexchange type coupling and hybridization between the TM, ligand I and Pb orbitals. 

    \item Fe and Ni polarize the nearby I significantly but with the same positive spin. 

    \item On the other hand, V, Co, and Cr shows the TM-I hybridization but with opposite spins. 

    \item Mn which results in virtually no spin polarization,its spin is highly localized on TM with a negligible extension of opposite spin in the bond. 
    This can be due to the weak anti-aligned polarization of bonding electrons (bond polarization) rather than a large ligand moment. 
    This correlates with d-orbital occupancy and covalency trends across 3d series. 
\end{enumerate}


 \begin{figure*}
    \includegraphics[scale=0.6]{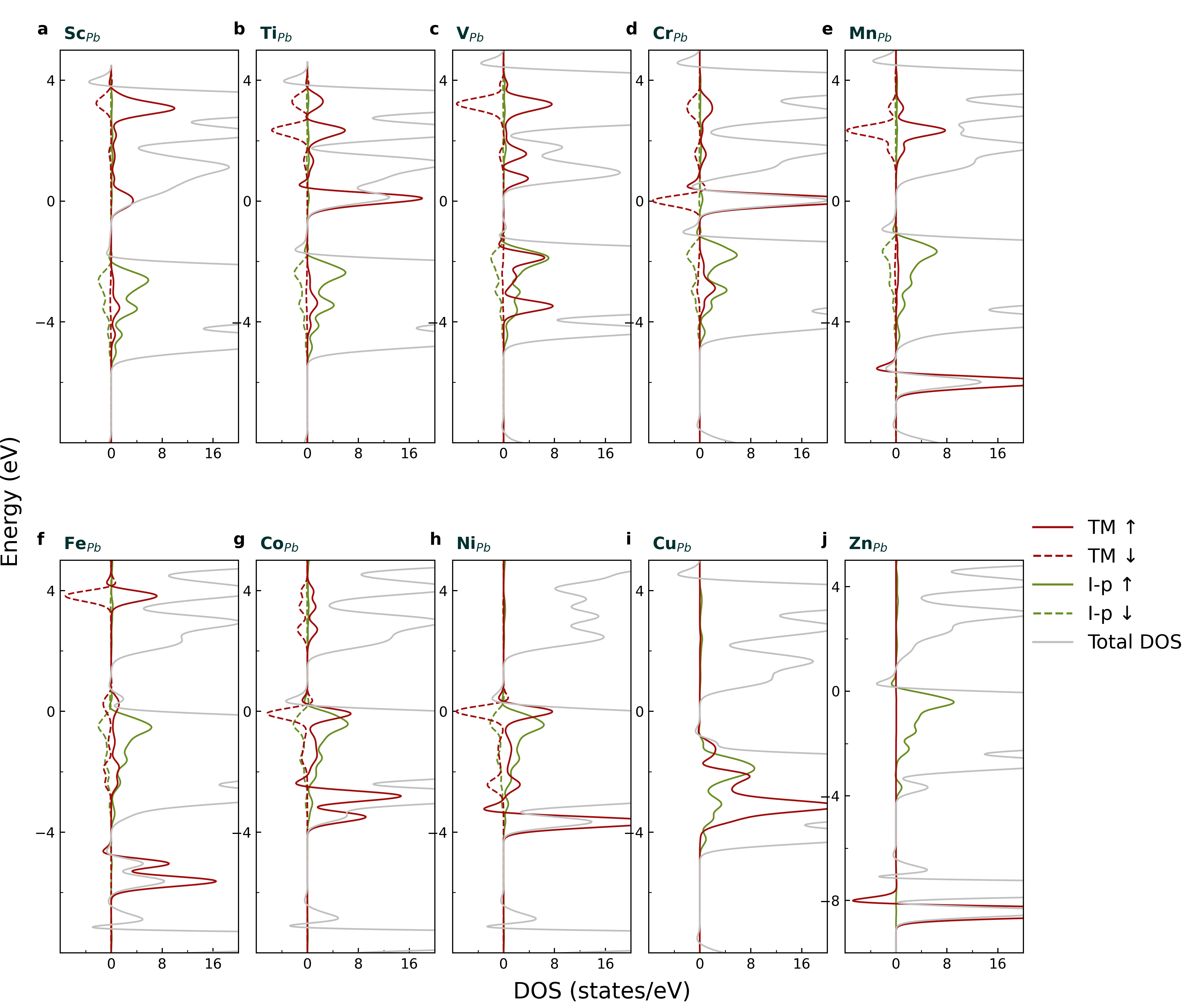}
    \caption{Spin polarized partial density of states.}
    \label{PDOS}
\end{figure*}

Confirmation of Iodine showing spin polarization and orbital overlap is obtained by the partial density of states (PDOS) in Fig. \ref{PDOS}. 
The dashed lined PDOS of Iodine is the evidence of negative spin polarization on Moving from Sc-Mn while leaving the Cr out for the moment, only majority spin states $\uparrow$ appear as occupied whereas for Fe, Co, and Ni, minority spin states $\downarrow$ can also be seen in valence band. 
For Sc and Ti, there are localized $\uparrow$ spin states exactly at Fermi energy with no host states which is common for mid gap states. This can be confirmed from Fig. \ref{DFEs} where only these dopants introduce deep states inside bandgap. 

In most systems, the d orbitals of dopants and 5p orbitals of ligand overlap at the top of valence band which is a sign of hybridization except Mn, Fe, and Zn which introduce d states around -5 eV, much lower than I-p states with almost no hybridization with I-p states. Only I-p states appear at Fermi level and at the band egde which means TMs do not introduce any detrimental state at the VB edge.


Fe, Co, Ni, and Zn doped systems become p-type (Fermi energy moves inside valence band), though Zn is nearly metallic. An interesting case is of Cr which becomes fully metallic. Cr-d states appear at or above the I-p orbitals closing the bandgap. 
Strong hybridization and more distorted structure destroys the band structure of the host (also the cause of more distrtion). 
Smaller or no difference between the energies of TM-d and I-p orbitals is responsible for strong hybridization, more overlap, more distortion and less stable defect.


    \section{\label{conclusion}Conclusion}

Overall, our results establish a unified picture in which local structural distortions, electronic hybridization, and spin polarization collectively determine the stability and electronic impact of TM substitution in \Iiii. 
The predominance of localized distortions and resonant defect states explains the robustness of this perovskite against a wide class of extrinsic defects, while also identifying specific dopants that are likely to be electronically benign or detrimental. 

    The distortions induced by TM substitution arise from strain-driven Van Vleck modes and are governed by a combination of ionic-radius mismatch, electronegativity differences, and TM-I orbital overlap. 
    Importantly, the distortion amplitude decays rapidly with distance from the defect site, indicating that the lattice responds locally rather than through long-range symmetry breaking. This localized accommodation underpins the structural stability and defect tolerance of \Iiii\ in the presence of substitutional defects.
    
    From an electronic-structure perspective, \Iiii\ exhibits a marked resistance to the formation of deep defect states. With the exception of Sc and Ti, most TM substitutions introduce resonant states that hybridize with the band edges rather than forming isolated mid-gap levels. While these resonant states can shift the band gap, they are substantially less detrimental to electronic transport than deep trap states, reinforcing the defect-tolerant character of the host material. Cr represents a limiting case, where strong TM-I hybridization closes the band gap and destabilizes the electronic structure, consistent with its enhanced structural distortion.

    Spin-polarized calculations further reveal that TM-I hybridization induces significant spin polarization on the ligands and, in some cases, on second-shell Pb atoms. The resulting patterns of spin leakage and cancellation reflect varying degrees of covalency and superexchange-like interactions across the 3d series. These trends correlate consistently with the partial density of states and provide an additional electronic signature of bonding character and defect stability.

    \putbib[references]
\end{bibunit}
    \clearpage

\begin{bibunit}[apsrev4-2]
\widetext
\setcounter{figure}{0}
\renewcommand{\thefigure}{S\arabic{figure}}
\setcounter{table}{0}
\renewcommand{\thetable}{S\arabic{table}}
\setcounter{page}{1}

\section*{\label{Supplementary Information}Supplementary Information\\
Defect Tolerance and Local Structural Response to 3d Transition-Metal Substitution in \Iiii}
\begin{center}
    Misbah Shaheen and Sheharyar Pervez*\\
Ghulam Ishaq Khan Institute of Engineering Sciences and Technology
\end{center}

\section{Methodology}

\subsection{Hubabrd Parameters}

Hubbard parameters used for transition metals in DFT calculations are given in the table \ref{hubbard} below:

\begin{table}[H]
\caption{\label{hubbard}Hubbard parameter values used in the study for transition metals. Most of the parameters are taken from materials project (MP) that were available there.}
\begin{ruledtabular}
\begin{tabular}{lrr}
 
 Material & Hubbard Parameter & Ref \\
         \hline

         Sc& 7 & \cite{Sc-Hubbard}\\
         \addlinespace
         Ti & 3.5 & \cite{Ti-Hubbard} \\

                                      \addlinespace
         V & 3.25 & \cite{MP} \\

                                      \addlinespace
         Cr & 3.7 & \cite{MP} \\

                                      \addlinespace
         Mn & 3.9 & \cite{MP}\\

                                      \addlinespace
         Fe & 5.3 & \cite{MP} \\

                                      \addlinespace
         Co  & 3.32 & \cite{MP} \\

                                      \addlinespace
        Ni & 6.2 & \cite{MP} \\

                                      \addlinespace
        Cu & 4.0 & \cite{Cu-Hubbard} \\

                                      \addlinespace
         Zn & 5.0 & \cite{Zn-Hubbard} \\

\end{tabular}
\end{ruledtabular}
\end{table}

\clearpage

\section{Results and Discussion}

\subsection{Stability Criteria}

\begin{figure}[h]
    \includegraphics[scale=0.50]{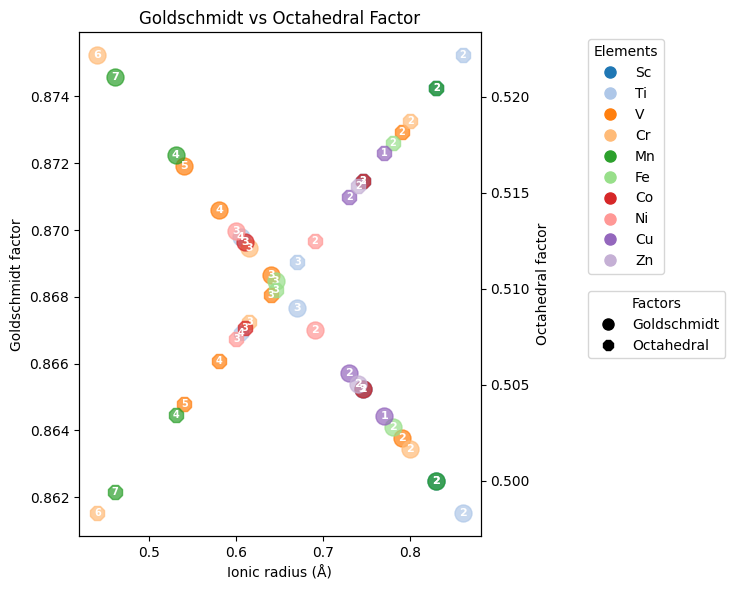}
    \caption{Goldschmidt tolerance(octahedral) factor of all systems showing the linearly increasing(decreasing) trend with the size of ionic radii. All the stable defects with their lowest oxidation states exist in the ionic radii range of 0.69-0.86 \AA.}
    \label{tmuvsradii}
\end{figure}

\subsection{Defect Formation Energies}
The DFEs and the other relevant obtained values from python code \textit{doped} are given in the tables \ref{table_Ipoor} and \ref{table_Irich}. 

\begin{table*}[h]
\caption{\label{table_Ipoor}Table showing the DFEs and all the other terms contributing to the DFEs in I-poor environment.}
\begin{tabular}{llrrrrrrrr}

\toprule
 &  & DeltaE & qEVBM & qEF & muRef & muFormal & Ecorr & Eform & DeltaEcorr \\
Defect & q &  &  &  &  &  &  &  &  \\
\midrule
\multirow[t]{4}{*}{VPbOhI3.15} & +3 & 9848.412000 & 9.730000 & 2.031000 & -9874.493000 & 0.000000 & 0 & -14.320000 & 0 \\
 & +2 & 9850.657000 & 6.487000 & 1.354000 & -9874.493000 & 0.000000 & 0 & -15.996000 & 0 \\
 & +1 & 9852.974000 & 3.243000 & 0.677000 & -9874.493000 & 0.000000 & 0 & -17.599000 & 0 \\
 & 0 & 9855.152000 & 0.000000 & 0.000000 & -9874.493000 & 0.000000 & 0 & -19.341000 & 0 \\
\cline{1-10}
\multirow[t]{2}{*}{CoPbOhI3.15} & +1 & 7768.629000 & 3.243000 & 0.677000 & -7785.833000 & 0.000000 & 0 & -13.284000 & 0 \\
 & 0 & 7771.046000 & 0.000000 & 0.000000 & -7785.833000 & 0.000000 & 0 & -14.787000 & 0 \\
\cline{1-10}
\multirow[t]{3}{*}{CrPbOhI3.15} & +4 & 9438.105000 & 12.973000 & 2.707000 & -9460.835000 & 0.000000 & 0 & -7.049000 & 0 \\
 & +1 & 9444.707000 & 3.243000 & 0.677000 & -9460.835000 & 0.000000 & 0 & -12.208000 & 0 \\
 & 0 & 9447.423000 & 0.000000 & 0.000000 & -9460.835000 & 0.000000 & 0 & -13.413000 & 0 \\
\cline{1-10}
\multirow[t]{2}{*}{CuPbOhI3.15} & 0 & 8930.506000 & 0.000000 & 0.000000 & -8945.839000 & 0.000000 & 0 & -15.333000 & 0 \\
 & -1 & 8933.028000 & -3.243000 & -0.677000 & -8945.839000 & 0.000000 & 0 & -16.731000 & 0 \\
\cline{1-10}
\multirow[t]{2}{*}{FePbOhI3.15} & +1 & 7349.931000 & 3.243000 & 0.677000 & -7368.122000 & 0.000000 & 0 & -14.270000 & 0 \\
 & 0 & 7350.917000 & 0.000000 & 0.000000 & -7368.122000 & 0.000000 & 0 & -17.205000 & 0 \\
\cline{1-10}
\multirow[t]{3}{*}{MnPbOhI3.15} & +5 & 8945.990000 & 16.217000 & 3.384000 & -8974.872000 & 0.000000 & 0 & -9.282000 & 0 \\
 & +2 & 8949.972000 & 6.487000 & 1.354000 & -8974.872000 & 0.000000 & 0 & -17.060000 & 0 \\
 & 0 & 8954.537000 & 0.000000 & 0.000000 & -8974.872000 & 0.000000 & 0 & -20.335000 & 0 \\
\cline{1-10}
\multirow[t]{2}{*}{NiPbOhI3.15} & +1 & 7158.272000 & 3.243000 & 0.677000 & -7177.465000 & 0.000000 & 0 & -15.272000 & 0 \\
 & 0 & 7160.666000 & 0.000000 & 0.000000 & -7177.465000 & 0.000000 & 0 & -16.799000 & 0 \\
\cline{1-10}
\multirow[t]{2}{*}{ScPbOhI3.15} & +1 & 10575.238000 & 3.243000 & 0.677000 & -10599.351000 & 1.670000 & 0 & -18.523000 & 0 \\
 & 0 & 10578.996000 & 0.000000 & 0.000000 & -10599.351000 & 1.670000 & 0 & -18.685000 & 0 \\
\cline{1-10}
\multirow[t]{3}{*}{TiPbOhI3.15} & +2 & 10201.936000 & 6.487000 & 1.354000 & -10225.417000 & 0.133000 & 0 & -15.508000 & 0 \\
 & +1 & 10204.746000 & 3.243000 & 0.677000 & -10225.417000 & 0.133000 & 0 & -16.618000 & 0 \\
 & 0 & 10208.622000 & 0.000000 & 0.000000 & -10225.417000 & 0.133000 & 0 & -16.662000 & 0 \\
\cline{1-10}
ZnPbOhI3.15 & 0 & 5548.256000 & 0.000000 & 0.000000 & -5564.893000 & 0.000000 & 0 & -16.637000 & 0 \\
\cline{1-10}
\bottomrule

\end{tabular}
\end{table*}
\clearpage
-----

\begin{table*}
\caption{\label{table_Irich}Table showing the DFEs and all the other terms contributing to the DFEs in I-rich environment.}
\begin{tabular}{llrrrrrrrr}
\toprule
 &  & DeltaE & qEVBM & qEF & muRef & muFormal & Ecorr & Eform & DeltaEcorr \\
Defect & q &  &  &  &  &  &  &  &  \\
\midrule
\multirow[t]{4}{*}{VPbOhI3.15} & +3 & 9848.412000 & 9.730000 & 2.031000 & -9874.493000 & -0.540000 & 0 & -14.860000 & 0 \\
 & +2 & 9850.657000 & 6.487000 & 1.354000 & -9874.493000 & -0.540000 & 0 & -16.536000 & 0 \\
 & +1 & 9852.974000 & 3.243000 & 0.677000 & -9874.493000 & -0.540000 & 0 & -18.138000 & 0 \\
 & 0 & 9855.152000 & 0.000000 & 0.000000 & -9874.493000 & -0.540000 & 0 & -19.880000 & 0 \\
\cline{1-10}
\multirow[t]{2}{*}{CoPbOhI3.15} & +1 & 7768.629000 & 3.243000 & 0.677000 & -7785.833000 & -1.829000 & 0 & -15.113000 & 0 \\
 & 0 & 7771.046000 & 0.000000 & 0.000000 & -7785.833000 & -1.829000 & 0 & -16.616000 & 0 \\
\cline{1-10}
\multirow[t]{3}{*}{CrPbOhI3.15} & +4 & 9438.105000 & 12.973000 & 2.707000 & -9460.835000 & -0.322000 & 0 & -7.371000 & 0 \\
 & +1 & 9444.707000 & 3.243000 & 0.677000 & -9460.835000 & -0.322000 & 0 & -12.530000 & 0 \\
 & 0 & 9447.423000 & 0.000000 & 0.000000 & -9460.835000 & -0.322000 & 0 & -13.734000 & 0 \\
\cline{1-10}
\multirow[t]{2}{*}{CuPbOhI3.15} & 0 & 8930.506000 & 0.000000 & 0.000000 & -8945.839000 & -1.627000 & 0 & -16.960000 & 0 \\
 & -1 & 8933.028000 & -3.243000 & -0.677000 & -8945.839000 & -1.627000 & 0 & -18.358000 & 0 \\
\cline{1-10}
\multirow[t]{2}{*}{FePbOhI3.15} & +1 & 7349.931000 & 3.243000 & 0.677000 & -7368.122000 & -1.131000 & 0 & -15.401000 & 0 \\
 & 0 & 7350.917000 & 0.000000 & 0.000000 & -7368.122000 & -1.131000 & 0 & -18.336000 & 0 \\
\cline{1-10}
\multirow[t]{3}{*}{MnPbOhI3.15} & +5 & 8945.990000 & 16.217000 & 3.384000 & -8974.872000 & -0.629000 & 0 & -9.911000 & 0 \\
 & +2 & 8949.972000 & 6.487000 & 1.354000 & -8974.872000 & -0.629000 & 0 & -17.689000 & 0 \\
 & 0 & 8954.537000 & 0.000000 & 0.000000 & -8974.872000 & -0.629000 & 0 & -20.964000 & 0 \\
\cline{1-10}
\multirow[t]{2}{*}{NiPbOhI3.15} & +1 & 7158.272000 & 3.243000 & 0.677000 & -7177.465000 & -1.622000 & 0 & -16.895000 & 0 \\
 & 0 & 7160.666000 & 0.000000 & 0.000000 & -7177.465000 & -1.622000 & 0 & -18.422000 & 0 \\
\cline{1-10}
\multirow[t]{2}{*}{ScPbOhI3.15} & +1 & 10575.238000 & 3.243000 & 0.677000 & -10599.351000 & 2.680000 & 0 & -17.513000 & 0 \\
 & 0 & 10578.996000 & 0.000000 & 0.000000 & -10599.351000 & 2.680000 & 0 & -17.675000 & 0 \\
\cline{1-10}
\multirow[t]{3}{*}{TiPbOhI3.15} & +2 & 10201.936000 & 6.487000 & 1.354000 & -10225.417000 & 1.620000 & 0 & -14.020000 & 0 \\
 & +1 & 10204.746000 & 3.243000 & 0.677000 & -10225.417000 & 1.620000 & 0 & -15.130000 & 0 \\
 & 0 & 10208.622000 & 0.000000 & 0.000000 & -10225.417000 & 1.620000 & 0 & -15.174000 & 0 \\
\cline{1-10}
ZnPbOhI3.15 & 0 & 5548.256000 & 0.000000 & 0.000000 & -5564.893000 & -0.098000 & 0 & -16.735000 & 0 \\
\cline{1-10}
\bottomrule

\end{tabular}

\end{table*}

\clearpage

\begin{table*}[h]
\centering
\caption{Formation energies per formula unit ($\Delta Ef$) of \ce{CsPbI3} and all competing phases, with k-meshes used in calculations. Only the lowest energy polymorphs are included.}
\label{tab:competingphaseformationenergies}
\begin{tabular}{cccc}
\hline
Formula & Space Group & E${\textrm{Hull}}$ (eV/atom) & $\Delta Ef$ (eV/fu) \\ \hline 
\ce{CsPbI3} & Pnma & 0.000 & -6.329 \\ 
\ce{Cs} & I$\overline{4}$3m & 0.000 & 0.000 \\ 
\ce{Sc} & P6${3}$/mmc & 0.000 & 0.000 \\ 
\ce{Ti} & P6/mmm & 0.000 & 0.000 \\ 
\ce{V} & Im$\overline{3}$m & 0.000 & 0.000 \\ 
\ce{Cr} & Im$\overline{3}$m & 0.000 & 0.000 \\ 
\ce{Mn} & I$\overline{4}$3m & 0.000 & 0.000 \\ 
\ce{Fe} & Im$\overline{3}$m & 0.000 & 0.000 \\ 
\ce{Co} & P6${3}$/mmc & 0.000 & 0.000 \\ 
\ce{Ni} & Fm$\overline{3}$m & 0.000 & 0.000 \\ 
\ce{Cu} & Fm$\overline{3}$m & 0.000 & 0.000 \\ 
\ce{Zn} & P6${3}$/mmc & 0.000 & 0.000 \\ 
\ce{Pb} & Fm$\overline{3}$m & 0.000 & 0.000 \\ 
\ce{I} & Cmce & 0.000 & 0.000 \\ 
\ce{CsI3} & Pnma & 0.000 & -4.467 \\ 
\ce{ScI3} & P6${3}$/mmc & 0.000 & -5.894 \\ 
\ce{TiI3} & Pmmn & 0.000 & -4.356 \\ 
\ce{VI2} & P$\overline{3}$m1 & 0.000 & -2.224 \\ 
\ce{CrI3} & P3${112}$ & 0.000 & -2.892 \\ 
\ce{NiI2} & R$\overline{3}$m & 0.000 & -1.141 \\ 
\ce{CuI} & P3m1 & 0.000 & -0.686 \\ 
\ce{PbI2} & P6${3}$mc & 0.000 & -2.763 \\ 
\ce{Cs2TiI6} & Fm$\overline{3}$m & 0.000 & -12.416 \\ 
\ce{Cs3Cr2I9} & P6${3}$/mmc & 0.000 & -16.408 \\ 
\ce{Cs3MnI5} & I4/mcm & 0.000 & -12.831 \\ 
\ce{CsFeI4} & P2${1}$/c & 0.000 & -5.649 \\ 
\ce{Cs3CoI5} & Pnma & 0.000 & -11.631 \\ 
\ce{Cs3Cu2I5} & Pnma & 0.000 & -12.035 \\ 
\ce{Cs2ZnI4} & Pna2${1}$ & 0.000 & -9.796 \\ 
\ce{Cs4PbI6} & R$\overline{3}$c & 0.000 & -16.868 \\ 
\hline
\end{tabular}
\end{table*}

\clearpage

\begin{figure*}[h]
    \includegraphics[scale=0.50]{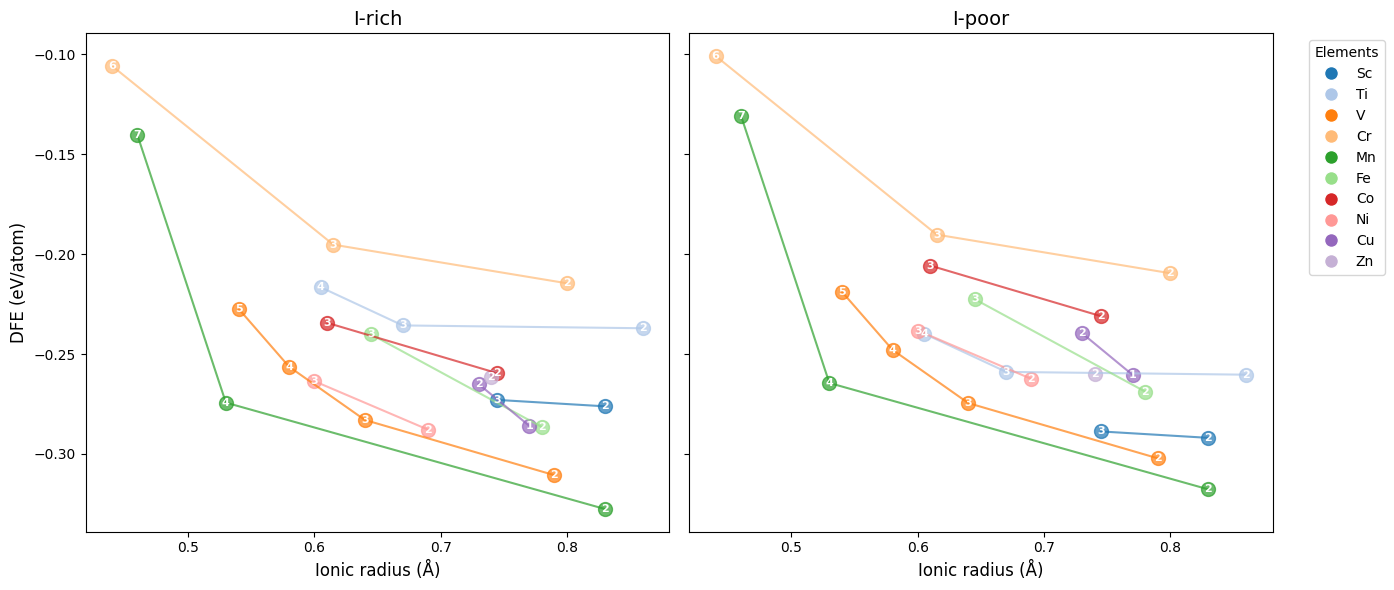}
    \caption{DFE vs ionic radii in both I-rich and poor regions showing the DFE trend moving toward stability with decreasing oxidation state.}
    \label{DFEvsionicradii}
\end{figure*}

Fig. \ref{DFEvsionicradii} shows the trend of formation energies w.r.t ionic radii which suggest the kind of exponential decay in DFEs with increasing oxidation states. Also the lowest oxidation states of each defect creates most stable defect.

\clearpage

\pagebreak

\subsection{Stability Analysis}

\begin{figure*}[h]
    \includegraphics[scale=0.70]{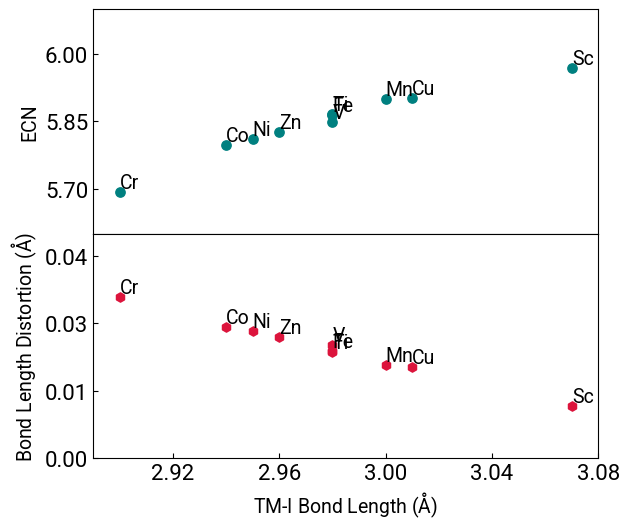}
    \caption{Distortion parameters: (top) Effective coordiantion number and (bottom) Bond length distortion.}
    \label{distparams}
\end{figure*}

Table \ref{bondlengths} shows that the TM-I bondlengths reduced in comaprison with the Pb-I bondlength in pristine \Iiii\ which in result elongate the Pb-I bondlength adjacent to TM-I (with I common) and introduce local distortions in structure. These local distortions are given in Fig. \ref{distparams} where effective coordination number (ECN) and bond length distortion shows the directl relation with TM-I bondlength. 

\begin{table}[h]
\caption{\label{bondlengths}Octahedral bondlengths for pristine \Iiii\ and doped structure showing the reduced values after doping.}
\begin{ruledtabular}
\begin{tabular}{lrr}
 
 Material & Pb-Halide bondlength & Dopant-Halide bondlength \\
         \hline
         \Iiii & 3.15 & -\\
         \addlinespace
         $\ce{Sc_{Pb}}$& 3.23 &3.07 \\
                                              \addlinespace
         $\ce{Ti_{Pb}}$& 3.32 & 2.98 \\

                                      \addlinespace
         $\ce{V_{Pb}}$& 3.32 & 2.98 \\

                                      \addlinespace
         $\ce{Cr_{Pb}}$& 3.40 & 2.90 \\

                                      \addlinespace
         $\ce{Mn_{Pb}}$& 3.30 & 3.00\\

                                      \addlinespace
         $\ce{Fe_{Pb}}$& 3.32 & 2.98 \\

                                      \addlinespace
         $\ce{Co_{Pb}}$ & 3.36& 2.94\\

                                      \addlinespace
        $\ce{Ni_{Pb}}$& 3.35 & 2.95 \\

                                      \addlinespace
        $\ce{Cu_{Pb}}$& 3.29 & 3.01 \\

                                      \addlinespace
         $\ce{Zn_{Pb}}$& 3.34 & 2.96 \\

\end{tabular}
\end{ruledtabular}
\end{table}


\clearpage

	\putbib[references]
\end{bibunit}

\end{document}